\documentclass[english,JIP]{ipsj}

\usepackage{graphicx}
\usepackage{latexsym}

\usepackage{url}
\usepackage{subcaption}
\usepackage{amsmath}

\usepackage[switch]{lineno}

\usepackage{xcolor}

\newcommand{\changed}[1]{{#1}}

\newcommand{\argmin}{\mathop{\rm arg~min}\limits}

\newtheorem{theorem}{Theorem}

\newtheorem{guideline}[theorem]{Guideline}

\newcounter{selectionguideline}

\def\Underline{\setbox0\hbox\bgroup\let\\\endUnderline}
\def\endUnderline{\vphantom{y}\egroup\smash{\underline{\box0}}\\}
\def\|{\verb|}

\received{2025}{4}{24}
\accepted{2025}{7}{3}

\usepackage[varg]{txfonts}%
\makeatletter%
\input{ot1txtt.fd}
\makeatother%

\begin{document}

\title{Selection Guidelines for Geo-Replicated SMR Protocols: A Communication Pattern-based Latency Modeling Approach}

\affiliate{TUTIMC}{Information and Media Center, Toyohashi University of Technology, Toyohashi, Aichi 441--8580, Japan}

\author{Kohya Shiozaki}{TUTCSE}[koya@dsl.cs.tut.ac.jp]
\author{Junya Nakamura}{TUTIMC}[junya@imc.tut.ac.jp]

\begin{abstract}
State machine replication (SMR) is a replication technique that ensures fault tolerance by duplicating a service.
Geo-replicated SMR is an enhanced version of SMR that distributes replicas in separate geographical locations, making the service more robust against large-scale disasters.
Several geo-replicated SMR protocols have been proposed in the literature, each tailored to specific requirements; for example, protocols designed to reduce latency by either sacrificing a part of their fault tolerance or limiting the content of responses to clients. 
However, this diversity complicates the decision-making process for selecting the best protocol for a particular service.
In this study, we introduce a latency estimation model for these SMR protocols based on the communication patterns of the protocols and perform simulations for various cases. 
Based on the simulation results and an experimental evaluation, we present five selection guidelines for geo-replicated SMR protocols based on their log management policy, distances between replicas, number of replicas, frequency of slow paths, and client distribution.
These selection guidelines enable determining the best geo-replicated SMR protocol for each situation.
\end{abstract}


\maketitle

\footnotetext{The preliminary version of this paper was published in Proceedings of the 26th International Symposium on Stabilization, Safety, and Security of Distributed Systems (SSS 2024)\cite{shiozaki_selection_2025}.}

\section{Introduction}
State machine replication (SMR)~\cite{schneider_implementing_1990,distler_byzantine_2021} is a technique that equips a service with fault tolerance by replicating the service across multiple servers called \emph{replicas}.
The replicas can reach a consensus on the processing order of the requests using the SMR protocol, thereby ensuring that the replicated service maintains a consistent state among all replicas.
Consequently, the service continues to operate even when several replicas fail.

Unlike conventional SMR, which places replicas within a single data center, geo-replicated SMR~\cite{coelho_geographic_2018,moraru_there_2013,yan_domino_2020, geopaxos+, Berger2022Aware, Xu2019ElasticRaft, Kostler2023Fluidity, Liu2017LeaderSet,  Enes2020Atlas, Eischer2020GuaranteedWrites} is a specialized form of SMR that distributes replicas across different geographical locations to enhance resilience against large-scale disasters.
Thus, the service can remain operational even if several replicas become inoperative because of major disasters such as earthquakes or tsunamis.

However, in geo-replicated SMR, the distributed placement of replicas significantly decreases responsiveness, which can be attributed to the increased and unbalanced communication delay resulting from more significant physical distances between replicas~\cite{numakura_evaluation_2019,Sousa2015,Chiba2023}.
In addition, the consensus of the SMR protocol among geographically distant replicas introduces an additional overhead. 
Therefore, a service designer needs to select a protocol carefully. 

Researchers have conducted numerous studies to address the responsiveness challenges of geo-replicated SMR.
Several geo-replicated SMR protocols have been introduced to enhance the responsiveness of geo-replicated SMR~\cite{coelho_geographic_2018,moraru_there_2013,yan_domino_2020, geopaxos+, Berger2022Aware, Xu2019ElasticRaft, Kostler2023Fluidity, Liu2017LeaderSet, Enes2020Atlas, Eischer2020GuaranteedWrites}.
Numakura et al.~\cite{numakura_evaluation_2019} proposed an optimization method for replica placement by modeling the latency of BFT-SMaRt protocol~\cite{bessani_state_2014}.

However, each protocol is optimized for different applications, which makes selecting the appropriate SMR protocol from among these developed protocols challenging.
The responsiveness of the protocols depends on the environment and requirements of the service. 
Further, the environment may change during service operations~\cite{numakura_evaluation_2019,Chiba2023}, thereby affecting the responsiveness.
While a comparative evaluation of SMR protocols under diverse conditions is essential, evaluating every possible combination of SMR protocols and replica placements is unrealistic.
In addition, to the best of our knowledge, no study provided a comprehensive overview of SMR protocol characteristics or selection guidelines for the protocols.
Therefore, selecting an SMR protocol that aligns with the service environment and requirements becomes highly challenging.

To address these issues, this study presents simulation-based selection guidelines for SMR protocols. 
These guidelines help service designers efficiently select an SMR protocol that is suited to their environment and requirements.
These guidelines are based on the communication patterns of five representative SMR protocols: MultiPaxos~\cite{van_renesse_paxos_2015}, Mencius~\cite{mao_mencius_2008}, FastPaxos~\cite{lamport_fast_2006}, Domino~\cite{yan_domino_2020}, and EPaxos~\cite{moraru_there_2013}.

To clarify the guidelines, we constructed a simulation model for each protocol to estimate the latency considering the communication time between clients and replicas. 
Using this model, we conducted simulations across various scenarios to reveal the distinct characteristics of each protocol and their effects on responsiveness. 
Further, we deployed an SMR on a cloud service to assess metrics that could not be assessed through simulation.
These wide-ranging measurements allowed us to evaluate the responsiveness of the protocols comprehensively and propose guidelines for selecting the most suitable protocol for each scenario.

The key contributions of this study are as follows:
\begin{enumerate}
    \item Establishing selection guidelines for geo-replicated SMR protocols.
    \item Proposing models that can estimate the latency of five representative SMR protocols accurately.
    \item Unveiling the detailed characteristics of these protocols.
\end{enumerate}
We can accurately predict the responsiveness of a service that incorporates geo-replicated SMR using the constructed models. 
Further, clarifying the characteristics of each protocol can contribute to the development of new protocols and related research.

The remainder of this paper is organized as follows.
Section~\ref{sec:Related_work} briefly describes geo-replicated SMR and existing research to capture the responsiveness of the SMR protocols.
Section~\ref{sec:Modeling} presents the latency models of five representative SMR protocols, and Section~\ref{sec:Guideline} proposes selection guidelines for SMR protocols based on the results of the simulations and experimental evaluation.
Finally, Section~\ref{sec:Conclusion} concludes the paper.

\section{Related Work}
\label{sec:Related_work}
SMR~\cite{schneider_implementing_1990,distler_byzantine_2021} is a technique that equips a service with fault tolerance by replicating across $n$ replicas to withstand up to $f$ replica failures. 
The replicas use an SMR protocol to reach a consensus on the processing order of the client requests. 
Once the replicas agree on the order, they \emph{commit} the commands requested by the clients to their replicated commit logs; each replica then independently executes the commands from the commit log in the same order to ensure consistency across all replicas.
\changed{In this way, SMR protocols guarantee strong consistency (specifically, linearizability~\cite{herlihy_linearizability_1990}) for write operations, ensuring that each replica maintains a consistent service state and allowing the service to continue operating even if several replicas fail.}

Replicas may experience \emph{crash} or \emph{Byzantine} faults during replication. 
In the former, a replica stops operating when it fails, and in the latter, a replica behaves arbitrarily when it fails. 
SMR is classified as a crash fault-tolerant SMR (CFT-SMR) or Byzantine fault-tolerant SMR (BFT-SMR) based on the assumed failure type.
The CFT-SMR can tolerate up to $f$ crash faults with $n \geq 2f + 1$ replicas, whereas the BFT-SMR can tolerate up to $f$ Byzantine faults with $n \geq 3f + 1$ replicas.
Many SMR~\cite{bessani_state_2014,van_renesse_paxos_2015,mao_mencius_2008,zhao_fast_2015,moraru_there_2013,yan_domino_2020} and geo-replicated SMR protocols~\cite{yan_domino_2020, coelho_geographic_2018, mao_mencius_2008, moraru_there_2013, tollman_epaxos_2021, geopaxos+} have been proposed in the literature.

A service designer must have sufficient knowledge of the characteristics of several protocols to select an appropriate protocol for geo-replicated SMR.
However, to the best of our knowledge, a comprehensive survey covering the various aspects of these protocols to help select an SMR protocol suited to the requirements of the designer remains lacking.
Although many studies that proposed SMR protocols compared their performances with those of existing protocols, their evaluations focused on demonstrating the superiority of the proposed protocols.
These studies did not provide this information, and therefore, guidelines for selecting geo-replicated SMR protocols are required.

Evaluating the performances of several SMR protocols in various scenarios is necessary to design such selection guidelines.
However, this evaluation is time-consuming, expensive, and impractical.
Therefore, several attempts have been made to capture the performance without actually building a replication.

For example, Castro~\cite{castro2001practical} presented a latency estimation model for the PBFT protocol considering the communication time between replicas and the processing delay within a replica.
Numakura et al.~\cite{numakura_evaluation_2019} modeled the latency of BFT-SMaRt~\cite{bessani_state_2014} and proposed an efficient method to determine the optimal replica placement in the geo-replicated SMR.
Inspired by Numakura et al.'s approach, we model and evaluate five representative CFT-SMR protocols in this paper.
\changed{Loruenser} et al.~\cite{loruenser_towards_2023} modeled PBFT~\cite{castro_practical_1999}; however, they focused on network layer parameters such as the packet loss rate of communication channels.
Thus, the attention layers differ from those in our approach.

Another approach involves emulating a wide-area network to measure the responsiveness of the SMR protocols~\cite{berger_does_2022,wang_tool_2022}.
This approach provides more detailed information than the modeling approach, while it requires more time and computational resources.

\section{SMR Protocol Latency Model}
\label{sec:Modeling}
This section constructs the latency estimation models for the five representative SMR protocols and verifies the estimation accuracy of the models.
We will use the models in the simulations in Section~\ref{sec:Guideline}.

\subsection{Model Construction}
We construct latency estimation models for the five representative SMR protocols (MultiPaxos~\cite{van_renesse_paxos_2015}, Mencius~\cite{mao_mencius_2008}, FastPaxos~\cite{lamport_fast_2006}, Domino~\cite{yan_domino_2020}, and EPaxos~\cite{moraru_there_2013}).
These protocols assume a partially synchronous network.
The modeling follows the approach proposed by Numakura et al.~\cite{numakura_evaluation_2019}, while the modeling target protocols differ from the paper. 
The latency is estimated by imitating the communication patterns during the consensus process of the SMR protocols.
For the estimation, we use the round-trip time (RTT) collected before estimation as the basic unit in the models.

First, to clarify the requirement regarding responses, we categorize services employing SMR into \emph{full-response} and \emph{status-response} services.
Each type exhibits different responsiveness characteristics.
In the full-response service, after committing a requested command, each replica executes the command to update its state and returns the execution result to the client.
In contrast, in the status-response service, each replica returns the success or failure of the command immediately after its commit, and then, it executes the command to update its state.

We aim to construct latency estimation models that can be used to compare responsiveness in various scenarios for formulating selection guidelines for geo-replicated SMR protocols, and therefore, hereafter, we construct latency models for the status-response services.
This is because, the command execution time, which is considered only in the full-response service, heavily varies depending on each service and its execution environment.
Therefore, we focus on the status-response service to simplify the models here and discuss the characteristics of the full-response service later in Section~\ref{sec:Full}.

We first introduce the latency estimation function $\mathit{EL}_{\mathit{avg}}(R,C)$ for a replica placement $R$ consisting of $n$ replicas and a set $C$ of client locations indicating the client distribution.
This function is the basis of the estimation and invokes the estimation model for each protocol.
$\mathit{EL}_{\mathit{avg}}(R, C)$ calculates the average latency of all clients by calculating the estimated latency of each client $c \in C$ as 
\begin{equation}
\label{eq:el_avg}
    \mathit{EL}_{\mathit{avg}}(R, C) = \frac{\sum_{c \in C}\mathit{EL}(R, c)}{| C |}
\end{equation}
In Equation \ref{eq:el_avg}, function $\mathit{EL}$ estimates the latency for each client based on the type of request as 
\begin{align}
\mathit{EL}(R, c) = & \mathit{Protocol}(R, \ell, c, p_{\mathit{slow}}) \times p_w \nonumber \\
                    & + \mathit{Read}(R, c) \times (1 - p_w) \label{EL_c}
\end{align}
where $p_w$ is the probability that a write request is sent; thus, the first and second terms of the right-hand side of the equation estimate the latency of write and read operations, respectively. 

In the first term, the function $\mathit{Protocol}$ takes four arguments $R$, $\ell$, $c$, and $p_\mathit{slow}$, where $\ell$ and $p_\mathit{slow}$ represent the leader replica and occurrence ratio of the slow path attributed to the concurrent client requests for the same data, respectively.
$\mathit{Protocol}$ calculates the latency of the write operation for each protocol based on its communication pattern, and we show the formulation of $\mathit{Protocol}$ in Sections~\ref{sec:multi-paxos-model}--\ref{sec:epaxos-model}.
Mencius and EPaxos omit the $\ell$ because they do not require a leader replica.
Similarly, MultiPaxos and Mencius omit $p_\mathit{slow}$ because no slow path occurs in the protocols.

In the second term, the function $\mathit{Read}(R, c)$ estimates the latency of read operations for each client $c$.
As read operations have a limited impact on the protocol selection, we mainly focus on write operations hereafter and will discuss read operations in Section~\ref{sec: read}.

\subsubsection{MultiPaxos}
\label{sec:multi-paxos-model}
\begin{figure}[tb]
\begin{center}
\includegraphics[scale=0.4]{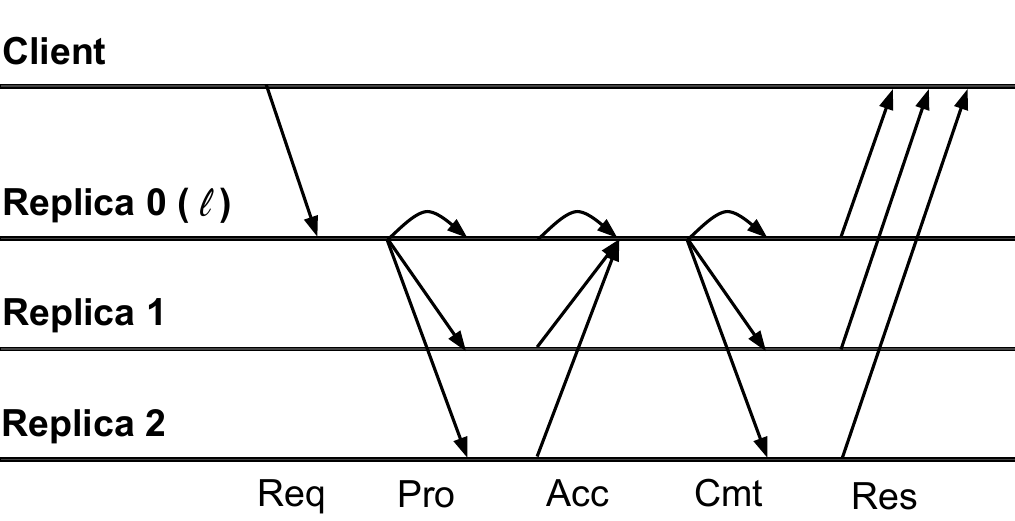}
\end{center}
\caption{Communication pattern of MultiPaxos for a write operation.}
\label{fig:multipaxos_communication}
\end{figure}

MultiPaxos~\cite{van_renesse_paxos_2015} extends the consensus protocol Paxos~\cite{paxos} to realize SMR.
MultiPaxos invokes individual Paxos instances for each request to agree on the execution order of its command.
Fig.~\ref{fig:multipaxos_communication} shows the communication pattern during a write operation in MultiPaxos.
This communication pattern is divided into the request, propose, accept, commit, and response phases. 
We calculate the timing for each replica when sending and receiving messages in each phase based on RTT, and we determine the latency of the write operation.
The elapsed times in each phase are denoted by $S_{\mathit{req}}$, $S_{\mathit{pro}}$, $S_{\mathit{acc}}$, $S_{\mathit{cmt}}$, and $S_{\mathit{res}}$. 
We add the superscript $S_{\mathit{pro}}^i$ to distinguish these times for a replica $r_i$.

Initially, the leader replica $\ell$ received a request from client $c$ at $S_\mathit{req}$, which is expressed as
\begin{equation}
    S_{\mathit{req}}=\mathit{OWD}(c, \ell)
\end{equation}
Here, $\mathit{OWD}(a, b)$ represents the communication time between two replicas or clients $a$ and $b$, which can be calculated from RTT between $a$ and $b$.
Upon reception, the leader replica $\ell$ transmits this request to all replicas as a propose message. 
The timing $S_{\mathit{pro}}^i$ at which each $r_i$ receives a propose message is
\begin{equation}
    S_{\mathit{pro}}^i=S_{\mathit{req}}+\mathit{OWD}(\ell, r_i)
\end{equation}
When $r_i$ receives the propose message, it replies with an accept message to the leader replica. 
The timing $S_{\mathit{acc}}$ at which the leader replica receives the majority of the accept messages is expressed as
\begin{equation}
    S_{\mathit{acc}}=\mathit{find}\left(T_{\mathit{acc}},\left\lceil \frac{n+1}{2} \right\rceil\right),
\end{equation}
where $T_{\mathit{acc}}=\{t \mid S_{\mathit{pro}}^i+\mathit{OWD}(r_i, \ell), 0 \leq i < n\}$ and the function $\mathit{find}(S, k)$ returns the $k$-th smallest element from set $S$.
Once the leader replica receives a sufficient number of accept messages, it sends the commit message to all replicas.
Consequently, the timing $S_{\mathit{cmt}}^i$ at which $r_i$ receives the commit message is
\begin{equation}
    S_{\mathit{cmt}}^i=S_{\mathit{acc}}+\mathit{OWD}(\ell,r_i)
\end{equation}
Lastly, once the commit message is received, the replica $r_i$ notifies the client of the successful commit via a response message. 
The client accepts the result upon receiving the first response message. 
Thus, the model $\mathit{MultiPaxos}(R, \ell, c)$ that estimates the latency of the write operation of MultiPaxos is given by
\begin{equation}
    \mathit{MultiPaxos}(R,\ell,c)=S_\mathit{res}=\min(T_\mathit{res}),
\end{equation}
where $T_{res}=\{t \mid S_{\mathit{cmt}}^i + \mathit{OWD}(r_i, c), 0 \leq i < n\}$.

\subsubsection{Mencius}
Mencius~\cite{mao_mencius_2008} was developed as an extension of MultiPaxos. 
Although it maintained the fundamental principles of MultiPaxos, it was tailored as an SMR protocol suited for wide area networks (WANs).
In MultiPaxos, a single fixed leader replica is responsible for managing all consensus instances, which causes the clients to send requests to the leader replica, thereby leading to poor responsiveness for clients geographically distant from the leader. 
To address this issue, Mencius partitions the commit log among all replicas in advance, thereby enabling them to manage consensus in parallel.
This enables clients to send requests to replicas that are geographically closer and have a lower RTT. 
Therefore, Mencius’s latency estimation model $\mathit{Mencius}(R, c)$ can be expressed using MultiPaxos model as
\begin{equation}
    \mathit{Mencius}(R,c)= \mathit{MultiPaxos}(R,\mathit{OptLeader}(R, c),c),
\end{equation}
where $\mathit{OptLeader}(R, c)=\argmin_{r \in R} \mathit{OWD}(c, r)$ is a function that returns replica $r \in R$ with the shortest communication time from client $c$.

\subsubsection{FastPaxos}
\begin{figure}[tb]
\begin{center}
\includegraphics[scale=0.4]{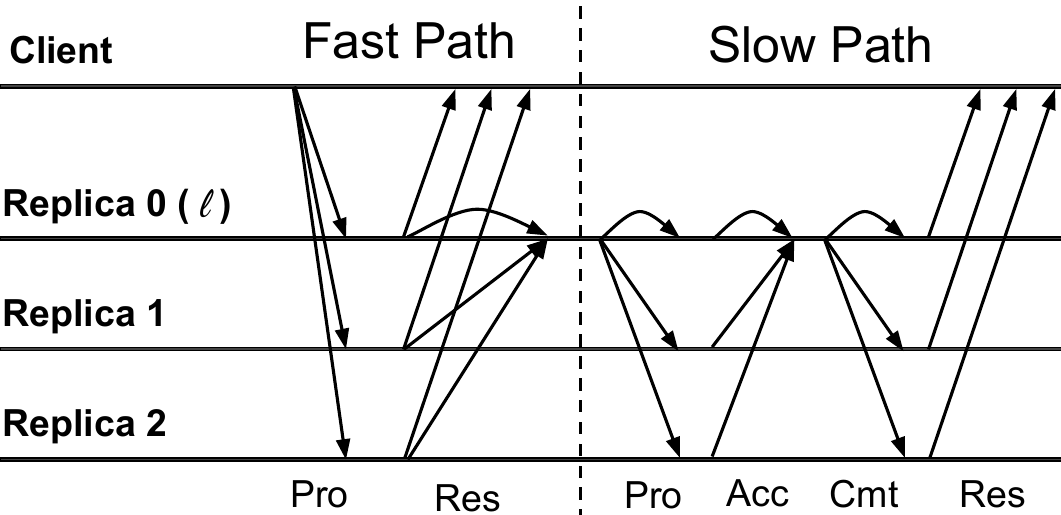}
\end{center}
\caption{Communication pattern of FastPaxos for a write operation.}
\label{fig:fastpaxos_communication}
\end{figure}

Fig.~\ref{fig:fastpaxos_communication} shows the communication pattern for the write operation in FastPaxos~\cite{lamport_fast_2006}, which has two consensus timings: the \emph{fast} and \emph{slow} paths.
In the fast path, client and replicas can quickly reach a consensus on the execution order of a command by proposing a request to all replicas directly without going through the leader.
However, if a request from another client for the same data arrives in the fast path (\emph{conflict}), the conflict is resolved in the slow path.
Thus, the latency of FastPaxos is estimated by summing the latencies of both the fast and slow paths while considering the occurrence rate $p_\mathit{slow}$ of the slow path.

First, replica $r_i$ receives a propose message from a client $c$ in the fast path at $S_{\mathit{fpro}}^i$ as
\begin{equation}
    S_{\mathit{fpro}}^i=\mathit{OWD}(c, r_i)
\end{equation}
Upon receiving the propose message, replica $r_i$ sends a response message to the client and the leader replica.
The client accepts the results when it receives $3n/4$ response messages; thus, the latency estimation model of the fast path $\mathit{FastPath}(R, c)$ is given by
\begin{equation}
    \mathit{FastPath}(R,c)= \mathit{find}\left(T_{\mathit{fres}},\left \lceil \frac{3n}{4} \right\rceil\right),
\end{equation}
where $T_{fres}=\{t \mid S_{\mathit{fpro}}^i + \mathit{OWD}(r_i, c), 0 \leq i < n\}$.

There are two reasons for the occurrence of slow paths in FastPaxos:
\begin{enumerate}
    \item The leader replica receives multiple requests in response messages from $3n/4$ replicas.
    \item The leader replica cannot receive response messages from $3n/4$ replicas within a certain time frame.
\end{enumerate}
We assume that the differences between these two timings do not have a significant effect on the latency and that they do not distinguish between them.

When a slow path is triggered, the leader replica $\ell$ sends a propose message to each replica.
The timing $S_{\mathit{spro}}^i$ at which replica $r_i$ receives this propose message is given by
\begin{equation}
    S_{\mathit{spro}}^i=\mathit{OWD}(\ell, r_i)
\end{equation}
After acceptance, the leader replica sends a commit message to each replica.
Thus, the timing $S_{\mathit{cmt}}^i$ at which replica $r_i$ receives this commit message is given by
\begin{equation}
    S_{\mathit{acc}}=\mathit{find}\left(T_{\mathit{acc}}, \left\lceil \frac{n+1}{2} \right\rceil\right),
\end{equation}
where $T_{\mathit{acc}}=\{t \mid S_{\mathit{spro}}^i+\mathit{OWD}(r_i, \ell), 0 \leq i < n\}$.
After acceptance, the leader replica sends a commit message to each replica.
Thus, the timing $S_{\mathit{cmt}}^i$ at which replica $r_i$ receives this commit message is given by
\begin{equation}
    S_{\mathit{cmt}}^i=S_{\mathit{acc}}+\mathit{OWD}(\ell,r_i)
\end{equation}
On receiving the commit message, the replica sends a response message to client $c$, which accepts the result when it receives the first message.
Thus, the model for estimating the latency of the slow path $\mathit{SlowPath}(R, \ell, c)$ is:
\begin{equation}
    \mathit{SlowPath}(R,\ell,c)=S_{sres}=\min(T_{sres}),
\end{equation}
where $T_{sres}=\{t \mid S_{\mathit{cmt}}^i + \mathit{OWD}(r_i, c), 0 \leq i < n\}$.
Therefore, the latency estimation model for the write operation of FastPaxos is 
\begin{align}
    \mathit{FastPaxos}(R,\ell,c,p_{\mathit{slow}}) = & \mathit{FastPath}(R,c) \nonumber \\
    & + \mathit{SlowPath}(R,\ell,c) \times p_{\mathit{slow}}
\end{align}

\changed{
The occurrence rate $p_{\mathit{slow}}$ of the slow path varies depending on the deployment environment and workload conditions. 
The rate tends to increase with many concurrent clients or skewed access patterns due to a higher likelihood of conflicts. 
Additionally, a heavy load on a replica or long network delays may prevent the required responses from arriving before the fast path timeout and increase the rate. 
Therefore, to estimate $p_{\mathit{slow}}$, it is necessary to analyze runtime logs collected from replicas to identify the conflict frequencies and the number of fast path timeouts. 
In addition, system metrics like replica CPU utilization and latency may provide useful indicators for estimating $p_{\mathit{slow}}$, as they correlate with conditions that may cause fast path timeouts.
In this way, the approximate magnitude and trend of $p_{\mathit{slow}}$ can be estimated.
}

\subsubsection{Domino}
Domino~\cite{yan_domino_2020} integrates both Mencius and FastPaxos as core protocols.
In Domino, each client measures the network speed between replicas periodically and selects an appropriate protocol based on these measurements. 
Domino estimates the latencies for the fast paths of FastPaxos and MultiPaxos when each replica becomes a leader. 
The system selects Mencius when the latency for MultiPaxos is the shortest; otherwise, it selects FastPaxos. 
This mechanism provides Domino with the capability to dynamically adjust to different network scenarios, thereby enabling a fast consensus.

Consequently, the latency estimation model of Domino for the write operation, $\mathit{Domino}(R, \ell, c, p_{\mathit{slow}})$, is
\begin{equation}
    \mathit{Domino}(R, \ell, c, p_{\mathit{slow}}) = \mathit{Mencius}(R, c)
\end{equation}
if $M \leq \mathit{FastPaxos}(R, \ell, c, 0)$ and 
\begin{equation}
    \mathit{Domino}(R, \ell, c, p_{\mathit{slow}}) = \mathit{FastPaxos}(R, \ell, c, p_{\mathit{slow}})
\end{equation}
for the other cases, where $M = \min_{r \in R} \mathit{MultiPaxos}(R, r , c)$ is the minimal latency via MultiPaxos across all replicas.

\subsubsection{EPaxos}
\label{sec:epaxos-model}
EPaxos~\cite{moraru_there_2013} is an SMR protocol optimized for WANs. 
Clients in EPaxos send requests to geographically closer replicas that have shorter communication times, similar to that in Mencius.
However, unlike Mencius, EPaxos uses different commit log management methods for replicas.
In Mencius, the commit log is partitioned equally among all the replicas in advance, and therefore, each replica is responsible for managing its assigned slot.
In contrast, EPaxos allows each replica to manage its commit log asynchronously. 
Upon receiving a request, the replica takes on the role of a commit log manager and coordinates the consensus with other replicas using the same communication pattern as that of Mencius.
This approach may cause conflicts in the command execution order if multiple replicas process requests for the same data simultaneously.
EPaxos addresses such conflicts with a slow path, similar to FastPaxos.

Therefore, the latency of EPaxos is estimated by summing the latencies of both the Mencius and slow paths, while considering the occurrence rate of the slow path $p_\mathit{slow}$.
The latency estimation model of EPaxos for a write operation, i.e., $ \mathit{EPaxos}(R, c, p_\mathit{slow}) $, is
\begin{align}
    \mathit{EPaxos}(R, c, p_{\mathit{slow}}) = & \mathit{Mencius}(R, c) \nonumber \\
    & + \mathit{SlowPath}\left(R, \right. \nonumber \\
    & \quad \left. \mathit{OptLeader}(R, c), c\right) \times p_{\mathit{slow}}
\end{align}

\subsubsection{Read Operation}\label{sec: read}
\changed{
A client issues a read operation to retrieve (possibly a part of) the current state of a replicated service.
While all protocols considered in this study are designed to ensure strong consistency (specifically, linearizability~\cite{herlihy_linearizability_1990}) for write operations, the consistency guarantees required for read operations, along with their associated communication patterns, are determined by the service layer's design and implementation.
In cases where the service layer also requires linearizability for read operations, clients need to read the state from the leader replica or a quorum.
Conversely, if the service layer prioritizes responsiveness over strict consistency, it may adopt weaker consistency models such as eventual consistency~\cite{bailis_eventual_consistency_2013}, allowing clients to read from any single replica.
In this study, we focus on the protocol-level characteristics and exclude read consistency models from our evaluation, as they are determined by the service layer and fall outside the scope of protocol-level comparison.
}

\subsection{Model Validation}
\label{sec:Model-Validation}
We evaluate the validity of the proposed latency estimation models to confirm whether they are sufficiently accurate for discussing the selection guidelines presented in Section~\ref{sec:Guideline}.
For the evaluation, we compare the latency estimated by the models with the experimentally measured latency.
We investigate the accuracy under two distinct environments for both the normal path (including the fast path) and the slow path.
The former environment involves all protocols without any request conflicts, whereas the latter targets the slow paths of FastPaxos and EPaxos under the condition where two or more requests conflict\footnote{We omit the verification of Domino's slow path because the slow path is derived from FastPaxos.}.

For validation, we measure the latencies of these protocols using Microsoft Azure\footnote{\url{https://azure.microsoft.com}}.
Table~\ref{tab:regions} lists the regions considered as potential locations for placing replicas and clients.
We adopted Domino’s~\cite{yan_domino_2020} implementation\footnote{\url{https://github.com/xnyan/domino.git}} to construct the SMR.
We derived the $\mathit{OWD}$ between the regions from the average RTTs measured using the \verb,ping, command taken 60 s prior to the experiment.
\begin{table}[tb]
    \centering
    \caption{Microsoft Azure regions used in this study.}
    \begin{tabular}{ll}
    \hline
    City Name     & Region Name \\ \hline
    Paris         & \verb|francecentral| \\
    Iowa          & \verb|centralus| \\
    Toronto       & \verb|canadacentral| \\
    Seoul         & \verb|koreacentral| \\
    London        & \verb|uksouth| \\
    California    & \verb|westus| \\
    Victoria      & \verb|australiasoutheast| \\
    Gävle         & \verb|swedencentral| \\
    São Paulo     & \verb|brazilsouth| \\
    Tokyo         & \verb|japaneast| \\
    Singapore     & \verb|southeastasia| \\
    Virginia      & \verb|eastus| \\
    Chennai       & \verb|southindia| \\
    \hline
    \end{tabular}
    \label{tab:regions}
\end{table}

\subsubsection{Estimation Accuracy in a Non-Conflict Environment}
\label{sec:fast_path_exp}
We investigate the estimation accuracy of the five proposed latency models in an environment without request conflicts (slow paths). 
We deploy a client in Tokyo and the replicas in three to seven locations, i.e., $n = 3, 4, 5, 6, 7$, randomly selected from the 13 regions listed in Table \ref{tab:regions}.
For each $n$, we selected 100 replica placement patterns randomly, resulting in 500 placements, and measured the actual and estimated latencies.

Generally, configuring an even number of replicas is considered rare because even replicas do not increase the number of tolerable faulty replicas.
The number of tolerable faulty replicas in the fast paths of FastPaxos and Domino can be improved by using replicas; for example, the fast path tolerates one faulty replica when $n = 4$.
Moreover, Loruenser et al.~clarified that additional replicas can suppress leader elections due to packet loss~\cite{loruenser_towards_2023}.
Thus, we consider even replicas.

For the actual latency, the client sends write operations (i.e., $p_w = 100\%$) to the replicas repeatedly in a closed loop, where each request is sent only after receiving the response to the previous one, for 10 s.
All other parameters of the SMR protocols adopt their default values.
We assume a status-response service, and therefore, replicas return responses to the client as soon as they have completed the commit phase. 
\changed{
Under these settings, we measure the time from when a client sends a request until it receives a response.  
To reduce the impact of outliers, we exclude values beyond 1.5 times the interquartile range (IQR) and calculate the average using the remaining measurements.%
}
The estimated latency for the placement and protocol is calculated using $\mathit{EL}_{avg}(R, C)$, where $R$ represents the placement of three randomly chosen replicas, $C$ represents Tokyo, $p_w$ is 100\%, $p_\mathit{slow}$ is 0\%, and $\mathit{Protocol}$ represents the protocol being measured.

\begin{figure*}[!tb]
    \centering
    \begin{subfigure}[t]{0.3\textwidth}
        \centering
        \includegraphics[width=55mm]{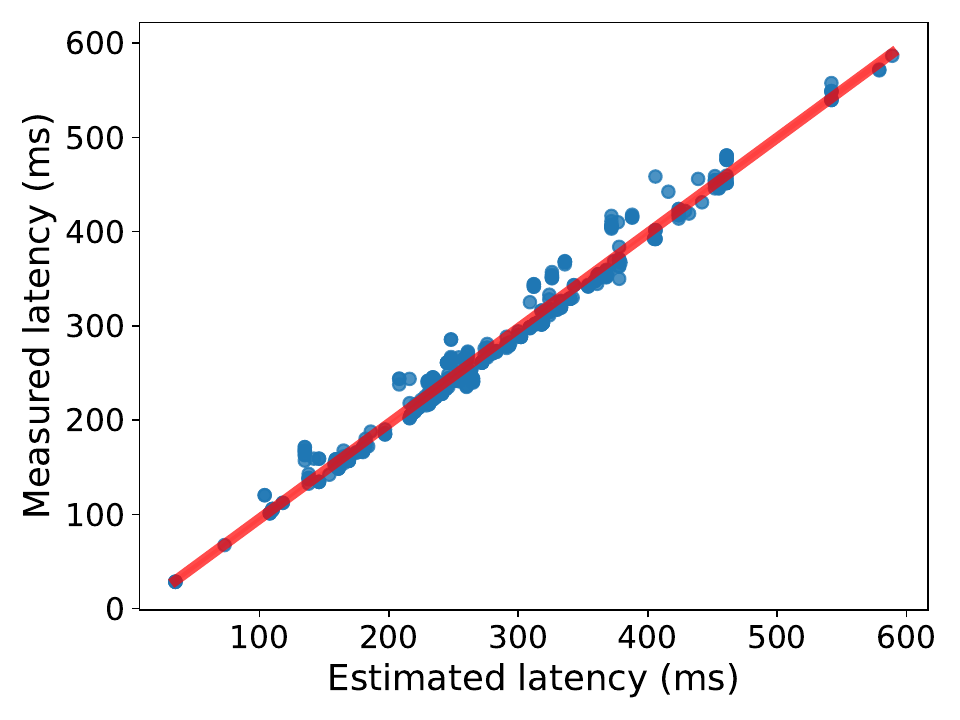}
        \caption{MultiPaxos}
    \end{subfigure}
    \hfill
    \begin{subfigure}[t]{0.3\textwidth}
        \centering
        \includegraphics[width=55mm]{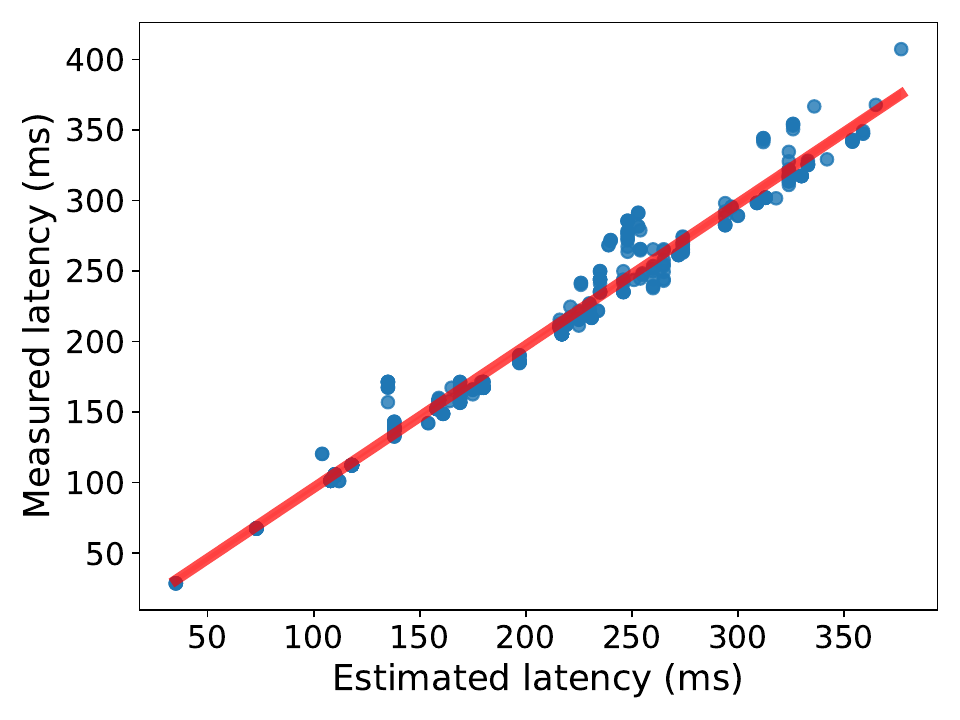}
        \caption{Mencius}
    \end{subfigure}
    \hfill
    \begin{subfigure}[t]{0.3\textwidth}
        \centering
        \includegraphics[width=55mm]{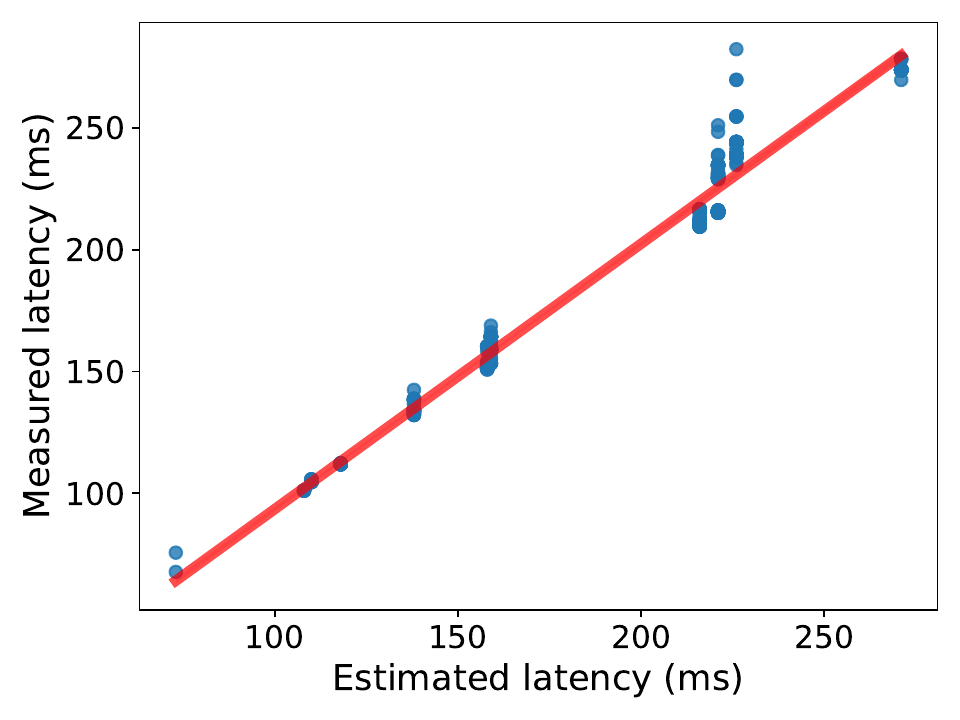}
        \caption{FastPaxos}
    \end{subfigure}
    \begin{subfigure}[t]{0.45\textwidth}
        \centering
        \includegraphics[width=55mm]{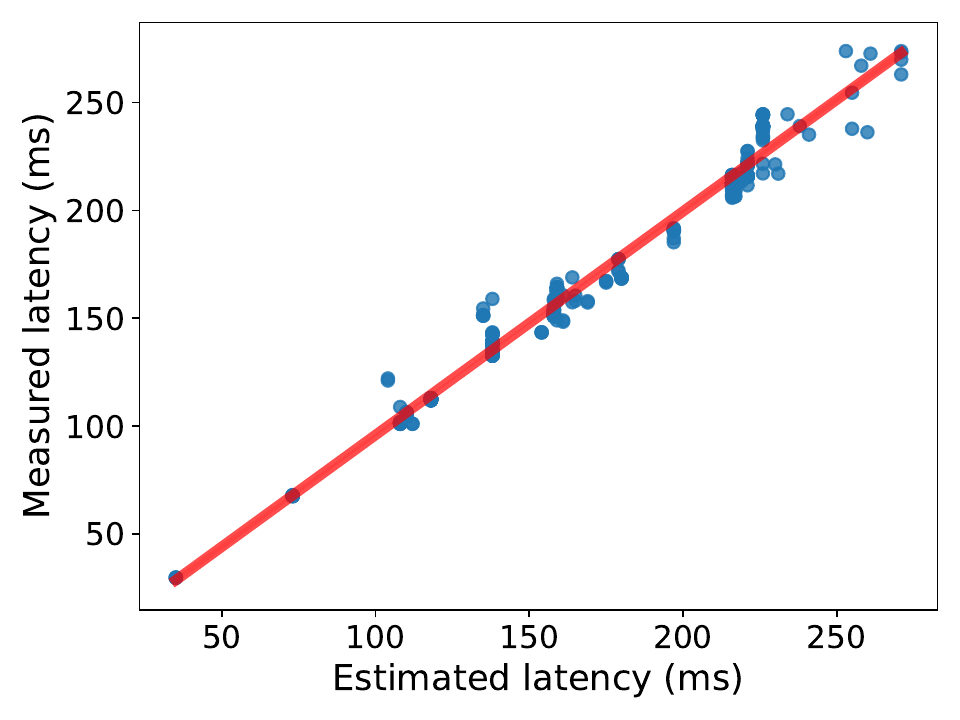}
        \caption{Domino}
    \end{subfigure}
    \hfill
    \begin{subfigure}[t]{0.45\textwidth}
        \centering
        \includegraphics[width=55mm]{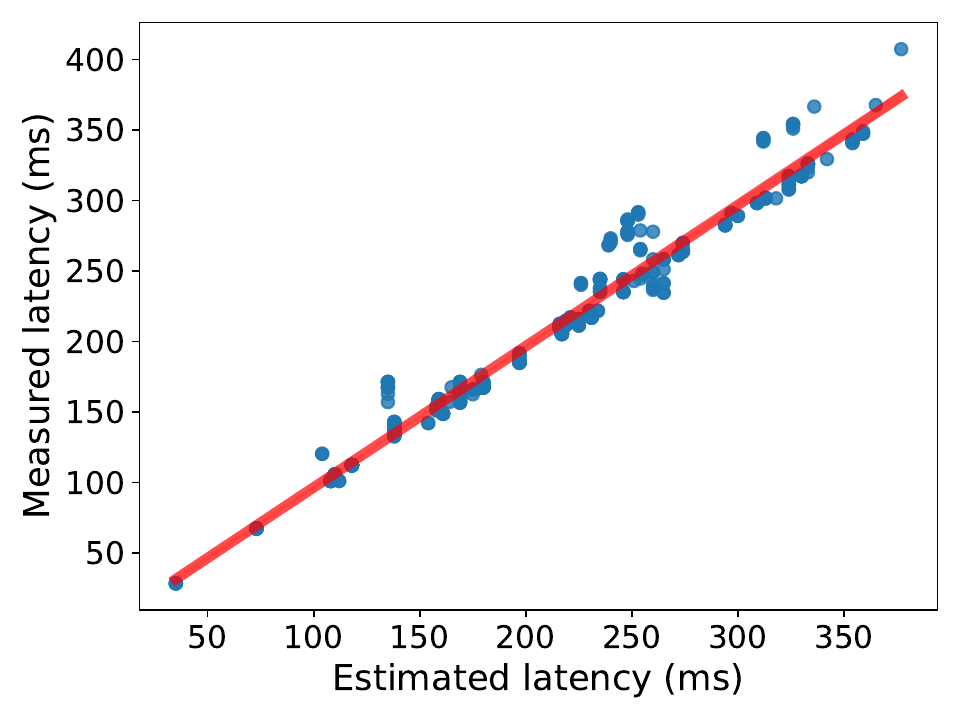}
        \caption{EPaxos}
    \end{subfigure}
    
    \caption{\changed{Estimated and measured latencies in a non-conflict environment}}
    \label{fig:latency-scatter-plot-non-conflict}
\end{figure*}

\changed{
Fig.~\ref{fig:latency-scatter-plot-non-conflict} shows scatter plots between the estimated and measured latencies for each protocol. 
Overall, the plots show a strong correlation between the estimated and actual values across all protocols, indicating that the estimation model performs well in non-conflict environments. 
In Fig.~\ref{fig:latency-scatter-plot-non-conflict}(c) for FastPaxos, fewer data points are shown; this is due to multiple placements producing identical or nearly identical latency values, resulting in overlapping points on the plot. 
Additionally, several instances in FastPaxos exhibit relatively large discrepancies between the estimated and measured latencies. 
The largest discrepancy was observed in one instance, where the estimated latency was 226\,ms while the measured latency was 282\,ms, resulting in a 56\,ms difference. 
This deviation is presumed to be caused by a temporary increase in RTT during the measurement of that instance.}

\changed{
Table~\ref{tab:model_accuracy_non_conflict} summarizes the correlation coefficients and root mean square error (RMSE) values for all protocols. 
The correlation coefficients are 0.932 or higher, and the RMSEs are 16.9\,ms or lower across the board, indicating that the proposed estimation model achieves high accuracy in non-conflict environments.
Moreover, the estimation for 1,000 placements took only 85\,s using a Python program running on a PC equipped with an Intel Core i5-12400 CPU, whereas actual latency measurement required 67,421\,s.
Based on these results, we can conclude that the estimation model allows us to rapidly and precisely estimate the latency of a placement.%
}

\begin{table*}[tb]%
    \centering
    \caption{\changed{Correlation coefficients and RMSE (ms) for different replica counts across protocols in a non-conflicting environment}}
    \label{tab:model_accuracy_non_conflict}
    \begin{tabular}{r|ccccc|ccccc}
        \hline
        & \multicolumn{5}{c|}{Correlation Coefficients} & \multicolumn{5}{c}{RMSE (ms)} \\
        \cline{2-11}
        Number of Replicas & MultiPaxos & Mencius & FastPaxos & Domino & EPaxos & MultiPaxos & Mencius & FastPaxos & Domino & EPaxos \\
        \hline
        3 & \changed{0.989} & 0.980 & 0.992 & \changed{0.989} & 0.979 & 15.6 & 16.7 & 7.3 & \changed{9.1} & 16.9 \\
        4 & \changed{0.991} & 0.987 & 0.995 & \changed{0.993} & 0.985 & \changed{12.6} & \changed{12.6} & \changed{5.4} & \changed{5.5} & \changed{13.3} \\
        5 & 0.988 & 0.990 & 0.987 & \changed{0.993} & 0.990 & 14.4 & 11.1 & \changed{7.9} & \changed{5.9} & 11.1 \\
        6 & \changed{0.991} & 0.988 & \changed{0.986} & \changed{0.991} & 0.983 & \changed{14.0} & 10.7 & 10.8 & \changed{6.7} & 12.5 \\
        7 & 0.991 & 0.993 & 0.932 & \changed{0.992} & 0.992 & 13.4 & \changed{8.5} & 13.0 & \changed{6.3} & \changed{9.3} \\
        \hline
        Total & \changed{0.990} & 0.987 & 0.985 & \changed{0.991} & 0.986 & \changed{14.0} & \changed{12.2} & 9.3 & \changed{6.8} & 12.9 \\
        \hline
    \end{tabular}
\end{table*}

\subsubsection{Estimation Accuracy in a Conflict Environment}
We investigate the estimation accuracy of the proposed latency models for FastPaxos and EPaxos in an environment in which request conflicts (slow paths) occur. 
The slow path is triggered when multiple clients simultaneously access the same data.
Therefore, we place three clients in the same location as the replicas and set $\alpha$, which represents the Zipfian distribution parameter, to 1000, such that conflicts occur frequently.
All other settings were consistent with those used in the previous experiment.
Under these settings, we measured the latency of the requests in the slow path, similar to that in Section~\ref{sec:fast_path_exp}.
The estimated latency for the replica placement and protocol was calculated using the value of $\mathit{EL}_{avg}(R, C)$, where $R$ and $C$ represent the placement of three randomly chosen replicas and same placement as the replicas, respectively; Further, $p_w$ is 100\%, $p_\mathit{slow}$ is 100\%, and $\mathit{Protocol}$ represents the protocol being measured.

\begin{figure*}[!tb]
    \centering
    \begin{subfigure}[t]{0.45\textwidth}
        \centering
        \includegraphics[width=55mm]{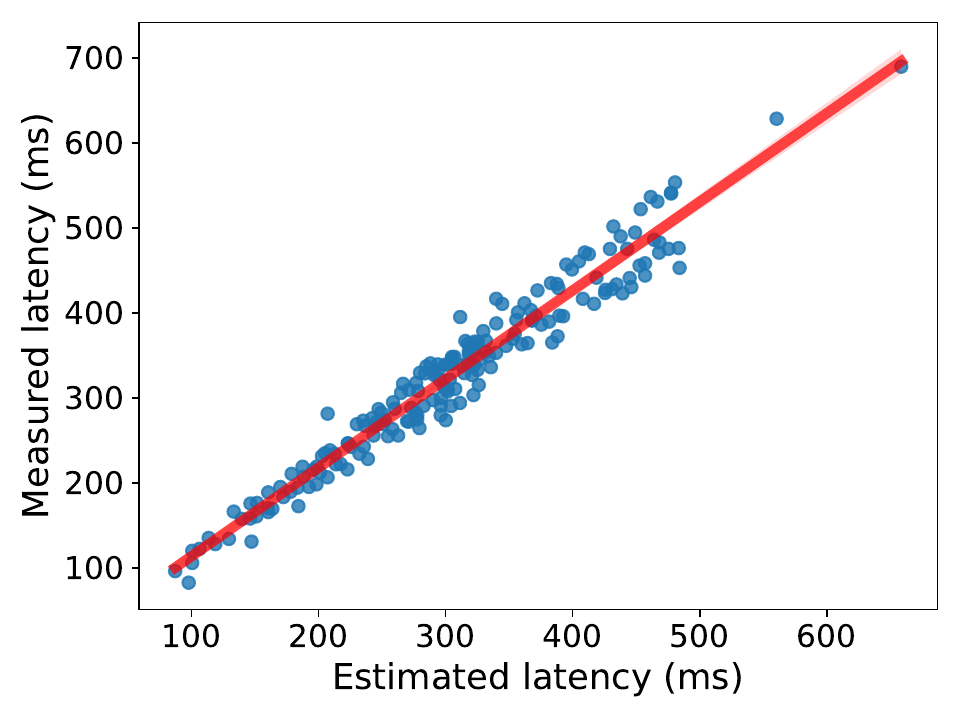}
        \caption{FastPaxos}
    \end{subfigure}
    \hfill
    \begin{subfigure}[t]{0.45\textwidth}
        \centering
        \includegraphics[width=55mm]{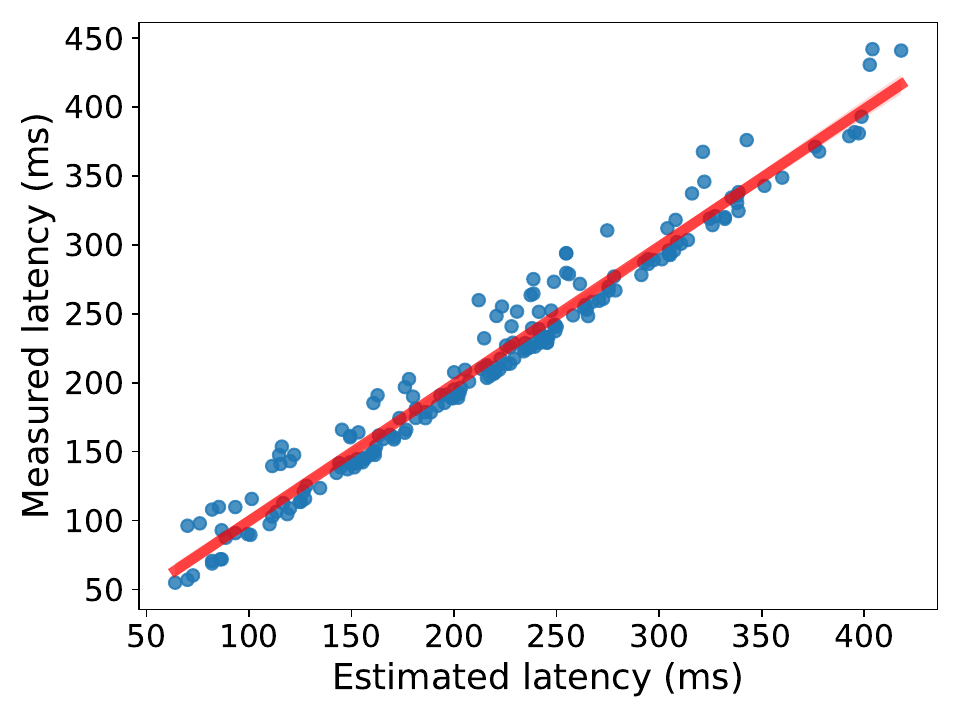}
        \caption{EPaxos}
    \end{subfigure}
    
    \caption{\changed{Estimated and measured latencies in a conflict environment}}
    \label{fig:latency-scatter-plot-conflict}
\end{figure*}

\begin{table}[tb]
    \centering
    \caption{\changed{Correlation coefficients and RMSE (ms) for different replica counts across protocols in a conflicting environment}}
    \label{tab:model_accuracy_conflict}
    \begin{tabular}{lcc}
        \hline
        Protocol & Correlation Coefficient & RMSE (ms)\\
        \hline
        FastPaxos  & \changed{0.977} & \changed{31.7}\\
        EPaxos     & 0.982 & 15.4\\
        \hline
    \end{tabular}
\end{table}
Figure \ref{fig:latency-scatter-plot-conflict} shows scatter plots between the estimated and measured latencies for each protocol, and Table~\ref{tab:model_accuracy_conflict} lists the correlation coefficients and RMSE.
Similar to the results in Section~\ref{sec:fast_path_exp}, the correlation coefficients for both protocols were \changed{0.977} or higher, and the RMSE of all protocols were \changed{31.7} ms or lower. 
The estimations were highly accurate also in conflicting environments.

\section{Selection Guidelines for Geo-Replicated SMR Protocols}
\label{sec:Guideline}
The guidelines for selecting geo-replicated SMR protocols are formulated as follows. 
In Section~\ref{sec:Status}, we investigated the characteristics of the protocols in the status-response service under various conditions using the estimation models.
In Section~\ref{sec:Full}, we analyzed the latency in the full-response service, considering the command execution time, through an experimental evaluation.
Finally, we present selection guidelines based on these investigation results in Section \ref{sec:Selection-Guidelines}.
Note that we present evaluation results in Sections~\ref{sec:Status} and~\ref{sec:Full} and discuss them in Section~\ref{sec:Selection-Guidelines}.

\subsection{Status-Response Service}
\label{sec:Status}
We simulated the responsiveness of SMR protocols under various conditions using the latency models. 
The simulation results aid in formulating the selection guidelines for the SMR protocols.

We used two client distributions, the \emph{global setting} and the \emph{US setting}, to clarify the effect of the geographical distribution tendencies of clients on responsiveness.
The global setting dispersed clients to Paris, São Paulo, Toronto, Victoria, and Tokyo, while the US setting concentrated clients in North America (Virginia, California, and Iowa).
We used the RTTs between regions published by Microsoft\footnote{\url{https://learn.microsoft.com/azure/networking/azure-network-latency}}.
We considered half of these values for $\mathit{OWD}$ in our latency model.

\changed{All simulations in this section are conducted as follows.
\begin{enumerate}
    \item Generate a set $\mathcal{R}_{\text{all}}$ of all possible combinations using $n$ regions selected from the 13 regions listed in Table~\ref{tab:regions}.
    \item Exclude from $\mathcal{R}_{\text{all}}$ any placement $R$ that contains two regions $r_i, r_j \in R$ such that the distance between them is less than $d$~km. Refer to the resulting set as $\mathcal{R}$.
    \item For each placement $R \in \mathcal{R}$, estimate protocol $P$'s latency $l^P_R$, given client distribution $C$, slow path occurrence rate $p_{\mathit{slow}}$, and write probability $p_w$.
    \item Use $\min \{ l^P_R \mid R \in \mathcal{R} \}$ as the latency of protocol $P$ under the given parameter configuration (i.e., $n$, $d$, $C$, $p_{\mathit{slow}}$, and $p_w$).
\end{enumerate}
Here, $P$ indicates the protocol that is being simulated, and we apply the above procedure to all five protocols to compute their respective latencies.
In all simulations, the write probability $p_w$ is fixed to 1.0, as this study focuses on write operations.  
Some other parameters (e.g., $n$, $d$, $C$, and $p_{\mathit{slow}}$) are varied across simulations, while the rest are fixed to the predefined values mentioned in each section. %
Hereafter, we analyze how each parameter affects protocol latency using the abovementioned procedure.}

\subsubsection{Effect of Geographical Distance Between Replicas}
The geo-replicated SMR distributes replicas to enhance resilience against large-scale disasters. 
Defining the disaster scale that the service must withstand is crucial when designing a service with a geo-replicated SMR, for example, by specifying the minimum geographical distance between the replicas.
However, assuming a large-scale disaster may increase the distance, thereby degrading the responsiveness.

\begin{figure*}[tb]
\centering
\begin{tabular}{ccc}
\begin{minipage}{0.32\textwidth}
\centering
\includegraphics[width=1.0\linewidth]{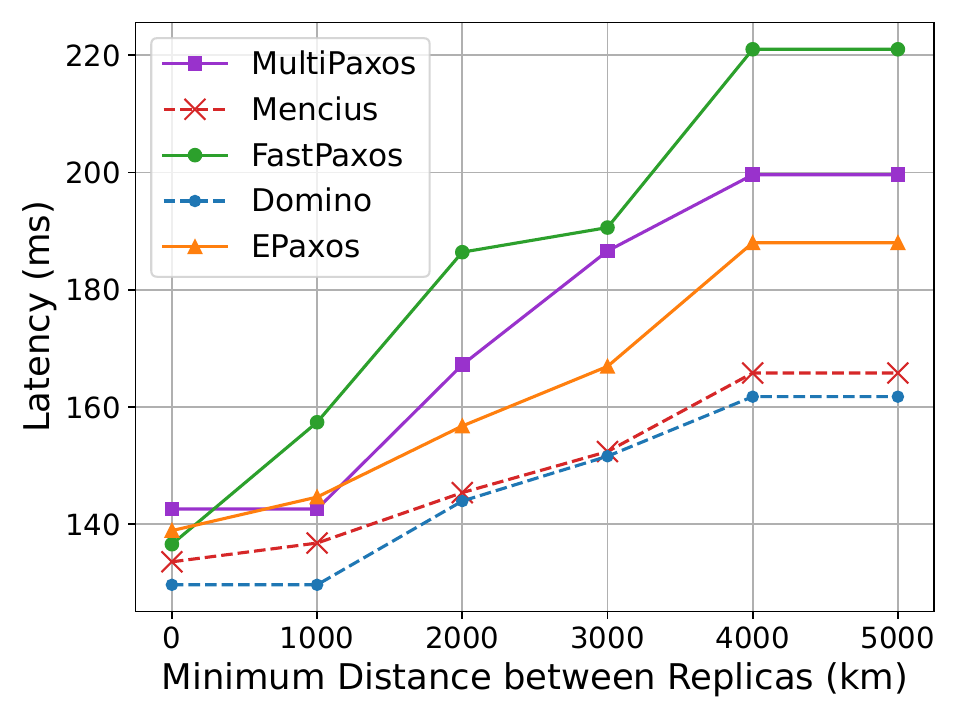}
\caption{\changed{Latency v.s. geographical distances of the replicas}}
\label{fig:latency_vs_distance}
\end{minipage}
&
\begin{minipage}{0.32\textwidth}
\centering
\includegraphics[width=1.0\linewidth]{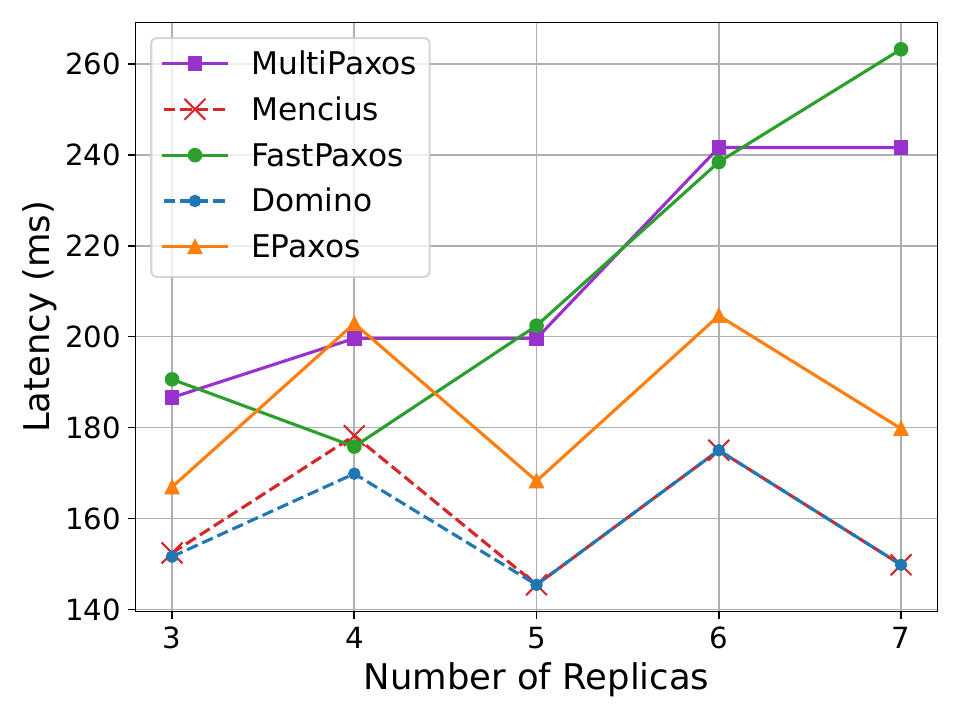}
\caption{\changed{Latency v.s. the number of replicas}}
\label{fig:latency_vs_replicas}
\end{minipage}
&
\begin{minipage}{0.32\textwidth}
\centering
\includegraphics[width=1.0\linewidth]{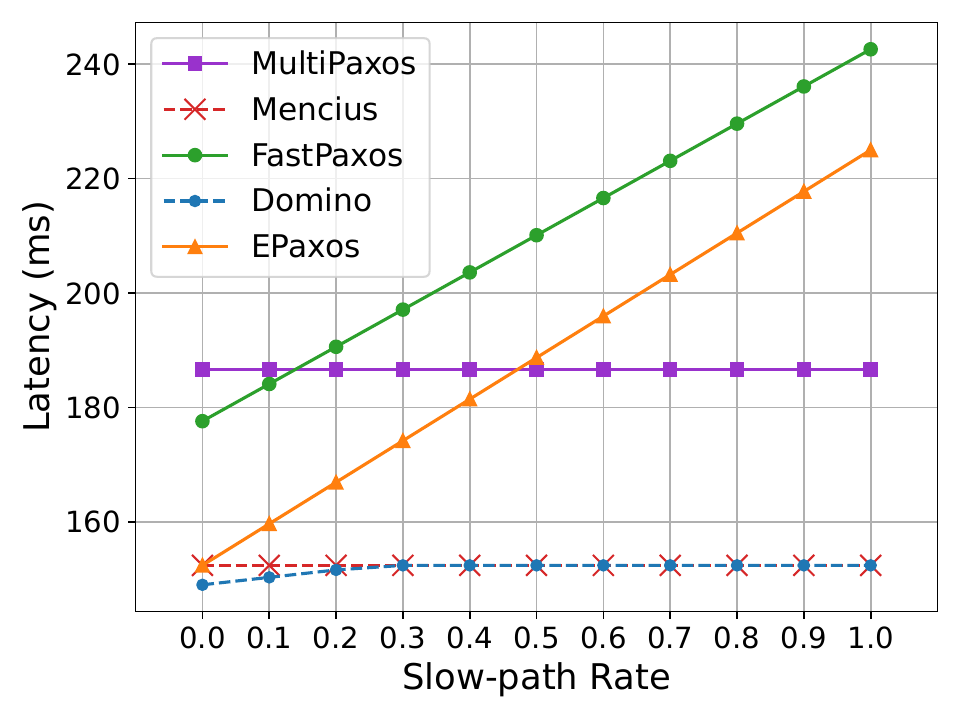}
\caption{\changed{Latency v.s. the slow-path occurrence rate}}
\label{fig:latency_vs_slow}
\end{minipage}
\end{tabular}
\end{figure*}

Given this background, we investigated the effect of the geographic distance between replicas on responsiveness. 
In the simulation, we varied the minimum distance $d$ between the replicas from 0 to 5000 km in increments of 1000 km.
We set the other parameters as follows: the number of replicas $n = 3$, slow-path occurrence rate $p_\mathit{slow} = 0.2$, and client distribution is the global setting. 
The simulation results are shown in Figure \ref{fig:latency_vs_distance}.
In this figure and the ones that follow, we show the results of Mencius and Domino with dotted lines to distinguish them from other protocols.
The reason for this will be discussed in section~\ref{sec:Full}

\subsubsection{Effect of the Number of Replicas}
Crucial parameters for designing an SMR service include the number of replicas, $n$, and the number of tolerable faulty replicas, $f$.
More replicas increase fault tolerance but involve more replicas in the consensus process, potentially decreasing responsiveness.
Therefore, we investigated the effect of the number of replicas on responsiveness.
We incrementally changed $n$ from three to seven in increments of one with the client distribution of the global setting, $d = 3000$\,km, and $p_\mathit{slow} = 0.2$.
The simulation results for all protocols are shown in Fig.~\ref{fig:latency_vs_replicas}.
Fig.~\ref{fig:latency_vs_replicas} indicates that the response time decreases with an increase in the number of replicas.
This trend differs from the expected behavior of typical SMR systems, where increasing the number of replicas leads to a larger quorum size, which increases the response time.
However, in the geo-replicated SMR, the placement of replicas can reduce response times despite the increase in the number of replicas.
When $n$ increases from three to four, the quorum size increases from two to three, thereby resulting in longer response times for Mencius.
However, when $n$ increases from four to five, the quorum size remains unchanged and the addition of an extra replica allows clients to form geographically closer quorums more flexibly, thereby resulting in shortened response times.

\subsubsection{Effect of the Slow-Path Occurrence Rate}
FastPaxos, EPaxos, and Domino trigger a slow path when multiple requests that manipulate the same data arise at the same time.
Additional communication is required when a slow path is triggered.
Consequently, in scenarios where a slow path occurs frequently, responsiveness may decline.
We investigated the effect of the slow-path occurrence rate on responsiveness.
We incrementally changed the slow-path occurrence rate $p_\mathit{slow}$ from zero to one in increments of 0.1 with the client distribution of the global setting, $n = 3$,  and $d = 3000$\ km. 
The simulation results are shown in Fig.~\ref{fig:latency_vs_slow}.

\begin{figure*}[tb]
\centering
\begin{tabular}{ccc}
\begin{minipage}{0.32\textwidth}
\centering
\includegraphics[width=1.0\linewidth]{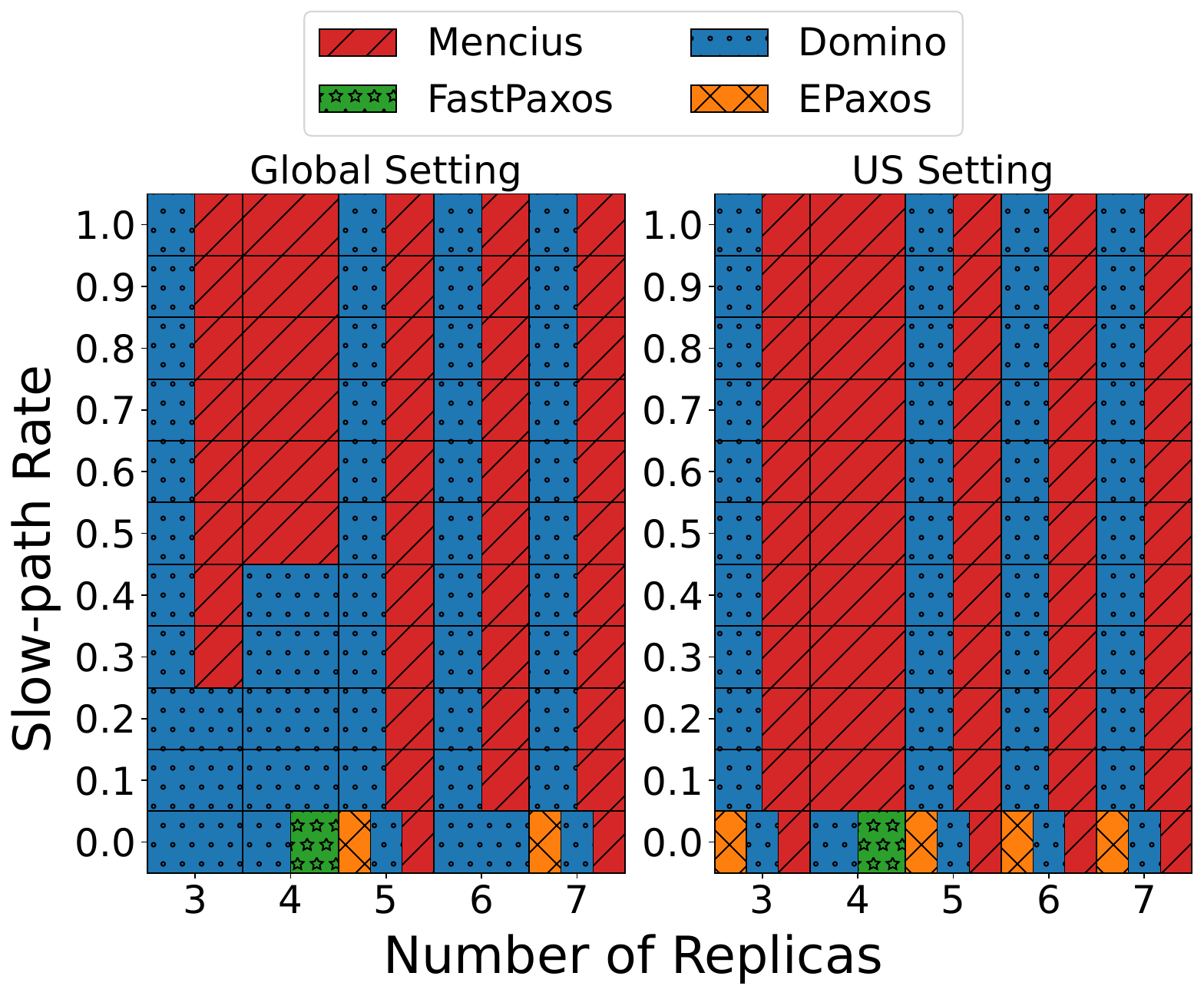}
\caption{\changed{Optimal protocol for each case for the status-response service}}
\label{fig:all_map}
\end{minipage}
&
\begin{minipage}{0.32\textwidth}
\centering
\includegraphics[width=1.0\linewidth]{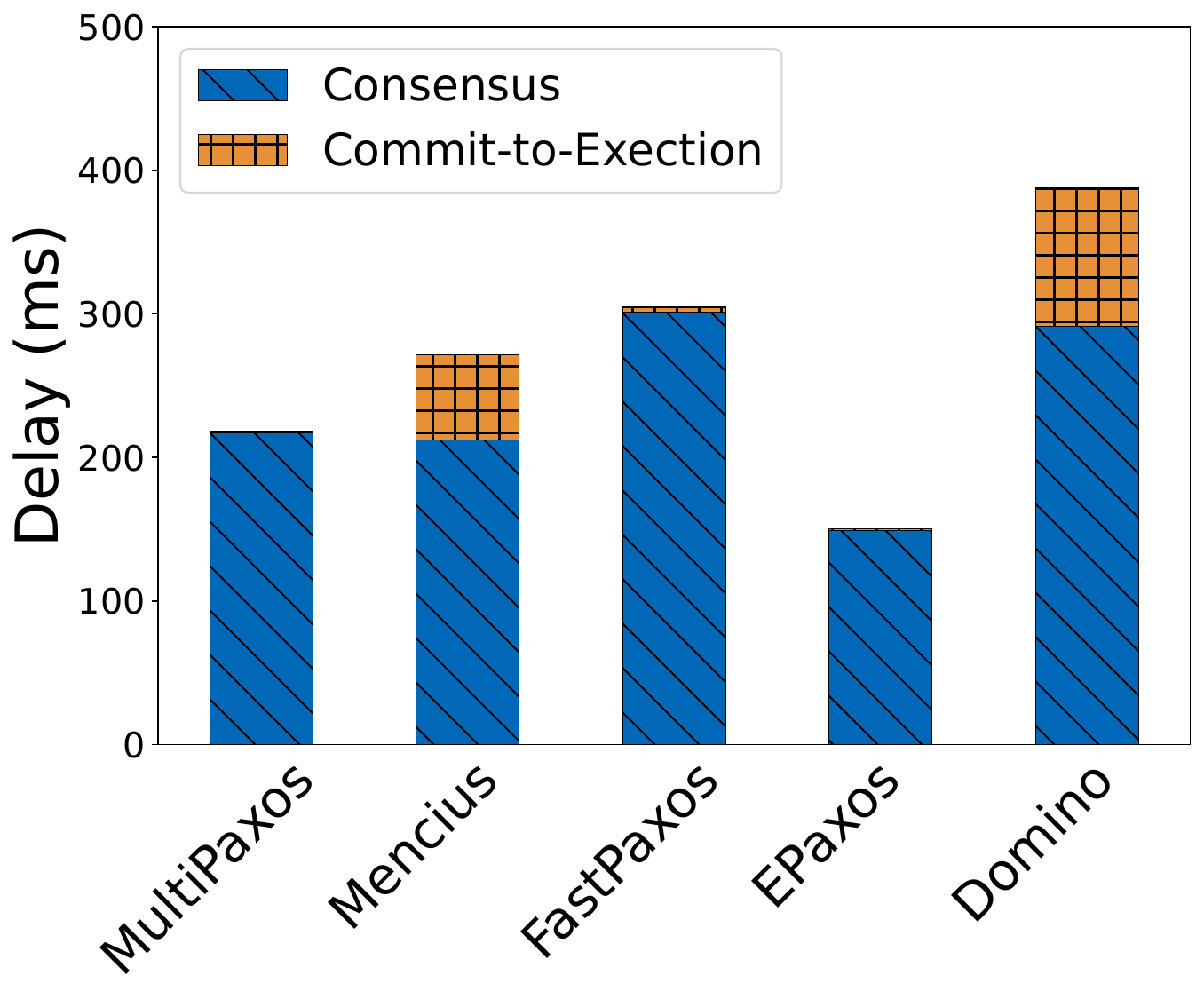}
\caption{\changed{Commit-to-execution delay of each protocol}}
\label{fig:stacked_bar}
\end{minipage}
&
\begin{minipage}{0.32\textwidth}
\centering
\includegraphics[width=1.0\linewidth]{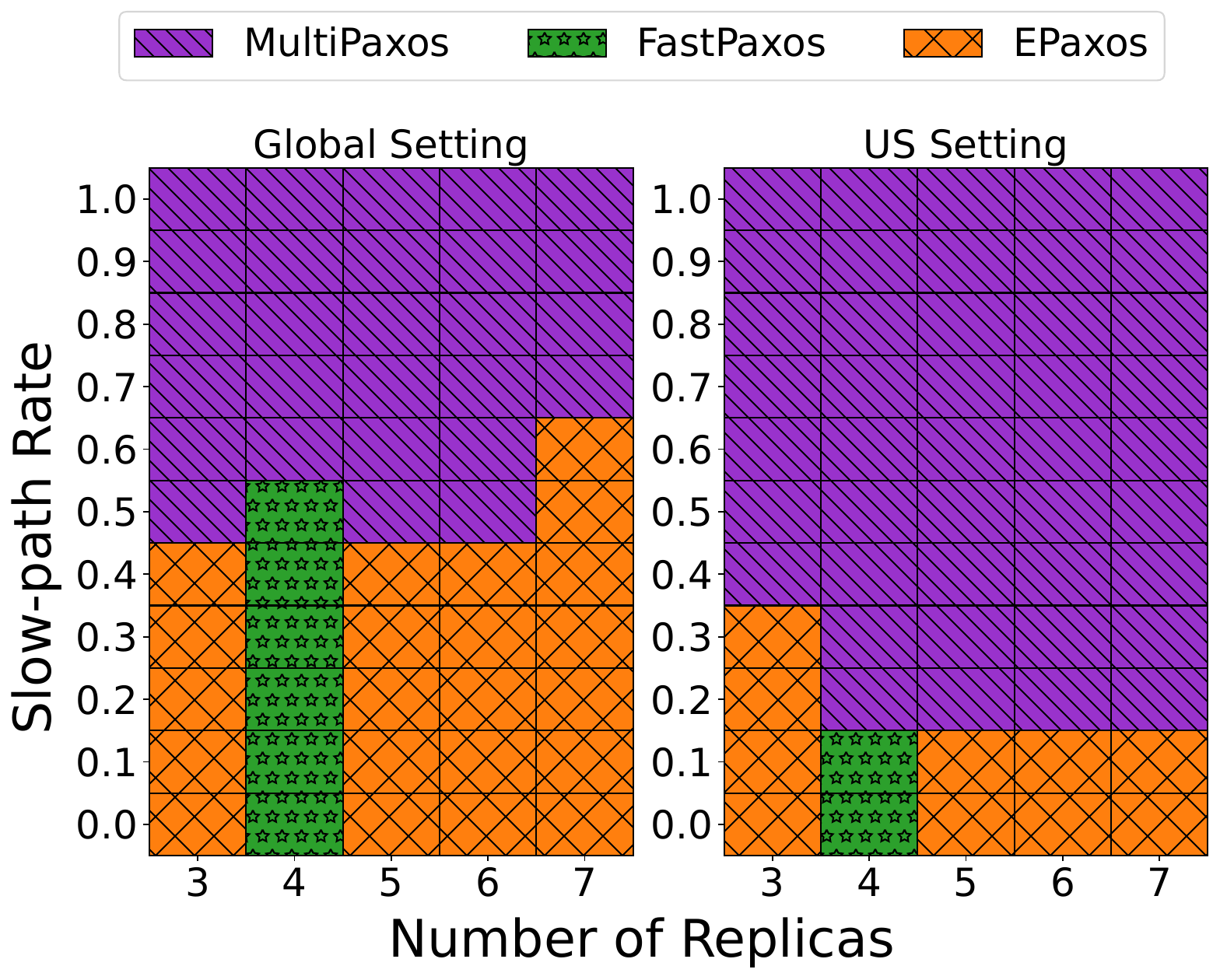}
\caption{\changed{Optimal protocol for each case of the full-response service}}
\label{fig:exec_map}
\end{minipage}
\end{tabular}
\end{figure*}

\subsubsection{Mapping of Optimal Protocols}
\label{sec:optimal-protocols-status-response}
Building on previous simulations, we explore how the optimal protocol for responsiveness varies by adjusting both the number of replicas $n$ and slow-path occurrence rate $p_\mathit{slow}$ under a distinct client distribution.
Given $d = 3000$\,km, we varied $n$ from three to seven and $p_\mathit{slow}$ from zero to one in increments of 0.1.
Fig.~\ref{fig:all_map} presents the optimal protocol for each combination of $n$ and $p_\mathit{slow}$.
In the figure, we fill a cell with two or more colors if multiple protocols are optimal for the cell.

In Fig.~\ref{fig:latency_vs_distance},~\ref{fig:latency_vs_replicas}, and~\ref{fig:latency_vs_slow}, Domino demonstrates lower latency compared to FastPaxos and Mencius. 
This is because Domino dynamically selects the appropriate protocol for each individual client.

\subsection{Full-Response Service}
\label{sec:Full}
In Section~\ref{sec:Modeling}, we modeled the latency of SMR protocols for the status-response service assuming that the execution time of commands remains unaffected by the protocol and evaluated them.
However, the evaluation does not consider the time from the commit of a command to the start of command execution, which we call the \emph{commit-to-execution delay}.
Although several protocols exhibit high performance in simulations of the status-response service, they may experience a performance drop when applied to the full-response service.

For example, in Mencius and Domino, the commit log management role is distributed equally among the replicas.
Replicas periodically share the state of the commit log they manage with others. 
Consequently, each replica obtains a full commit log and executes commands in the log from the oldest to update the state of the replicated service.
In this process, a replica cannot execute a command until all the commit logs preceding it are filled with a command or the special command, \emph{no-op}, thereby indicating the absence of a command.
Therefore, even if a command can be committed quickly, its execution may be delayed.

\subsubsection{Effect of the Commit-to-Execution Delay}
We conducted an experiment using an actual SMR implementation built on Microsoft Azure to measure the commit-to-execution delay.
For this experiment, we deployed three replicas in Tokyo, California, and London and three clients in Seoul, Toronto, and Paris.
All other settings were aligned with the experiments described in Section~\ref{sec:fast_path_exp}.
Given these settings, we measured the time from when a command was committed until its execution began at the replica.
The average duration across all requests, measured in the three replicas, is considered the commit-to-execution delay for each protocol with the client distribution.

The experimental results are presented in Fig.~\ref{fig:stacked_bar}. 
As indicated in the chart, the protocols were divided into two groups: 
The first group comprises  MultiPaxos, FastPaxos, and EPaxos.
Their commit-to-execution delay is small (0.82—3.11 ms), and therefore, these protocols are suitable for full-response services.
The delay of the second group (Mencius and Domino) is large (59.0--96.2 ms) and about 100 times longer than that of the first group.
This significant delay arises because the replicas in the second group must gather information from the commit log managed by other replicas to execute a command.

\subsubsection{Mapping of Optimal Protocols}
These results are used to re-evaluate the optimal protocol for the full-response service similar to that in Section~\ref{sec:optimal-protocols-status-response}.
Fig.~\ref{fig:exec_map} shows the corresponding results.
Although Mencius and Domino performed well in Section \ref{sec:Status}, this result indicates that these protocols are not suitable for full-response services due to their long commit-to-execution delay.
These aspects must be considered to clarify the selection guidelines for SMR protocols, which is why we distinguish Mencius and Domino from the others in Figs.~\ref{fig:latency_vs_distance}--\ref{fig:latency_vs_slow}.

\subsection{Selection Guidelines}
\label{sec:Selection-Guidelines}
Finally, we discuss the results from both the simulations and experiments, and formulate the selection guidelines for SMR protocols based on these findings.

We present the following guidelines for a full-response service based on the results shown in Fig.~\ref{fig:stacked_bar}:
\begin{guideline}
For a full-response service, selecting SMR protocols that do not have pre-partitioned commit logs (i.e., MultiPaxos, FastPaxos, EPaxos) can enhance responsiveness.
\end{guideline}

For the distance between replicas, Fig.~\ref{fig:latency_vs_distance} shows that the latency increases with an increase in the distance between replicas.
In addition, a larger inter-replica distance resulted in a more significant difference in the latency among protocols, which implies that protocol selection plays an important role, particularly when considering major disasters.
Among all protocols, Domino outperformed the others at most distances, with Mencius slightly surpassing Domino at 3000 km. 
Among the protocols suitable for the full-response service (i.e., MultiPaxos, FastPaxos, and EPaxos), MultiPaxos performs the best with a maximum inter-replica distance of up to 1000 km, whereas EPaxos performs better for greater distances. 
Based on these findings, we derived the following guideline:
\begin{guideline}
For the status-response service, Domino can improve responsiveness regardless of distance constraints. 
For a full-response service, MultiPaxos and EPaxos should be selected for shorter and larger inter-replica distances, respectively.
\end{guideline}

For the effect of the number of replicas on responsiveness shown in Fig.~\ref{fig:latency_vs_replicas}, the greater the number of replicas, the longer is the latency.
However, EPaxos, Mencius, and Domino can effectively reduce degradation because each client can send a request to the replica near it, not the leader.
Conversely, MultiPaxos and FastPaxos exhibited significant performance deterioration with an increasing number of replicas. Based on these observations, we propose the following guideline:
\begin{guideline}
For a service requiring a higher fault tolerance (i.e., more replicas), using protocols that prioritize geographically closer replicas can minimize performance degradation. 
Mencius and Domino are suitable for the status-response service, and EPaxos for the full-response service.
\end{guideline}

In scenarios where massive requests are sent to a small portion of the data, the slow path occurs frequently.
Fig.~\ref{fig:latency_vs_slow} shows that the latencies of EPaxos, FastPaxos, and Domino increase with an increase in the slow-path occurrence rate.
EPaxos and FastPaxos experience significant degradation, whereas the increase in latency for Domino is moderate.
This can be attributed to its hybrid approach, where clients can select a suitable protocol between FastPaxos and Mencius.
Based on these findings, we propose the following guidelines.
\begin{guideline}
In a service with high data access skew, where a small subset of data is frequently accessed by many clients simultaneously (leading to frequent slow paths), it would be advantageous to use protocols that avoid slow paths, such as MultiPaxos or Mencius.
This selection can help mitigate performance degradation caused by slow paths.
\end{guideline}

Client distribution affects replication responsiveness.
The simulation involving all protocols (Fig.~\ref{fig:all_map}) indicates that protocols that partition commit logs in advance, such as Domino and Mencius, are superior in most cases, regardless of the client distribution trends.
In the global setting (Fig.~\ref{fig:all_map}), Mencius is optimal for smaller values of $n$, whereas Domino is best for larger values of $n$ because Domino's clients choose FastPaxos and Mencius flexibly.
In the comparison of the protocols with a small commit-to-execution delay (Fig.~\ref{fig:exec_map}), MultiPaxos is selected as the optimal protocol in many cases.
This trend becomes stronger if the client distribution is dense (as in the US setting).
Based on these considerations, we derived the following guideline:
\begin{guideline}
If we can predict client distribution in advance, we can select an optimal protocol.
For status-response services, Mencius and Domino can improve responsiveness, regardless of client distribution.
For a full-response service, EPaxos is beneficial for a dispersed client distribution, whereas MultiPaxos is recommended when the clients are densely located.
\end{guideline}

\section{Conclusion}
\label{sec:Conclusion}
In this study, we proposed five selection guidelines for geo-replicated SMR protocols. 
We constructed latency estimation models for the consensus process of five representative CFT-SMR protocols and simulated their performance under various conditions to formulate these guidelines. 
Correlation coefficients between the latency estimated by the models and the actual latency were 0.932 or higher, and the RMSE was \changed{16.9} ms or lower, thereby demonstrating its high accuracy.
In addition, we implemented SMR on a cloud service to measure metrics that could not be evaluated through simulations. 
A service designer can easily select the best SMR protocol suited to their service requirements using these guidelines.
In addition, our models allowed us to predict the responsiveness of a geo-replicated SMR service with high precision.
The classification characteristics of these protocols are expected to contribute to the development of new geo-replicated SMR protocols.

We plan to expand our models to include BFT-SMR and other CFT-SMR protocols that are not considered here.
Further, we will develop selection guidelines that consider throughput, another crucial performance metric, in addition to latency.

\begin{acknowledgment}
This work was supported by JSPS KAKENHI Grant Numbers JP20KK0232 and JP22K11971.
\end{acknowledgment}

\bibliographystyle{ipsjsort-e}
\bibliography{references}

\begin{biography}
\profile[fig/kohya-shiozaki.jpg]{Kohya Shiozaki}{
received the B.E.~and M.E.~degrees from Toyohashi University of Technology, Japan, in 2023 and 2025, respectively.
He is currently a Ph.D.~student at Toyohashi University of Technology.
His research interests include distributed systems.
}
\profile[fig/nakamura-junya.jpeg]{Junya Nakamura}{
received the B.E.~and M.E.~degrees from Toyohashi University of Technology, Japan, in 2006 and 2008, respectively, and the Ph.D.~in Information Science and Technology from Osaka University in 2014.
He is currently an associate professor at Information and Media Center, Toyohashi University of Technology.
His research interests include theoretical and practical aspects of distributed algorithms and systems.
He is a member of ACM, IEEE, IEEE Computer Society, IEICE, and IPSJ.}
\end{biography}

\label{ipsj@lastpage}
\addtocounter{page}{1}
\end{document}